\documentclass[graybox]{svmult}

\usepackage{draftwatermark}
\usepackage{comment}
\usepackage{enumitem}
\usepackage{type1cm}        
\usepackage{makeidx}         
\usepackage{graphicx}        
\usepackage{multicol}        
\usepackage[bottom]{footmisc}

\usepackage{newtxtext}       %
\usepackage[varvw]{newtxmath}       

\makeindex             

\begin{document}

\title*{Edge Computing for IoT}
\author{Balqees Talal Hasan and Ali Kadhum Idrees}
\institute{Balqees Talal Hasan, \at Dept. of Computer and Information Engineering, Nineveh University, Mosul, Iraq , \email{balqees.hasan@uoninevah.edu.iq} \and Ali Kadhum Idrees, \at Dept. of Information Networks, University of Babylon, Babylon, Iraq, \email{ali.idrees@uobabylon.edu.iq}  }
%
%
\maketitle

\abstract {Over the past few years, The idea of edge computing has seen substantial expansion in both academic and industrial circles. This computing approach has garnered attention due to its integrating role in advancing various state-of-the-art technologies such as Internet of Things (IoT) , 5G, artificial intelligence, and augmented reality. In this chapter, we introduce computing paradigms for IoT, offering an overview of the current cutting-edge computing approaches that can be used with IoT. Furthermore, we go deeper into edge computing paradigms, specifically focusing on cloudlet and mobile edge computing. After that, we investigate the architecture of edge computing-based IoT, its advantages, and the technologies that make Edge computing-based IoT possible, including artificial intelligence and lightweight virtualization. Additionally, we review real-life case studies of how edge computing is applied in IoT-based Intelligent Systems, including areas like healthcare, manufacturing, agriculture, and transportation. Finally, we discuss current research obstacles and outline potential future directions for further investigation in this domain. }

\section{Introduction}
\label{sec:1}
In as early as 1966, an insightful prediction emerged from Karl Steinbuch, a pioneer in German computer science. He predicted that within a few decades, computers would be a necessary component of almost every industrial product.The term "pervasive computing" was first introduced by W. Mark in 1999. It means integrating computers into everyday objects so seamlessly that they become a natural and unnoticed part of the environment, with people interacting with them effortlessly. In the same year (1999), The term "Internet of Things" was coined by Kevin Ashton at a presentation at Procter \& Gamble (P\&G) \cite{cloudlet_2}.
\\ IoT is a new paradigm for attaching various physical objects to the Internet so they can interact and make informed decisions. The technologies that fall under this paradigm include pervasive computing, RFID,  communication technologies, sensor networks, and Internet protocols. In IoT, physical things have the ability to intelligently collaborate and establish connections with the Internet, operating autonomously and introducing innovative applications. These applications span a variety of industries, such as manufacturing, transportation,healthcare, industrial automation, and emergency response \cite{edge3,ali1}. 
\\ IoT has become permeated our daily lives, providing crucial measuring and data-gathering capabilities that influence our decision-making. Numerous sensors and gadgets run continuously, producing data and enabling vital communication over complex networks. It is challenging to execute complicated computations on most of the IoT devices due to their limited CPU and energy resources. In general, IoT devices collect data and transmit it to more robust processing centers for analysis\cite{ali3}.  The data is subjected to extra processing and analysis at these centers \cite{ali4}. In order to lighten the load on resources and avoid overuse of them, edge computing has become prevalent as a novel way to address IoT and local computing requirements \cite{edge2}. In edge computing, small servers that are located closer to users are used to process data. In close proximity to the devices of the consumers, these edge servers are capable of doing complex tasks and storing enormous volumes of data. As a result of their proximity to users, processing and storage at the network edge becomes faster and more efficient \cite{advantages}. In 2022, the worldwide edge computing market was worth USD 11.24 billion. Experts predict that it will experience significant expansion, with an expected yearly From 2023 to 2030, the growth rate is 37.9\% \cite{edge_1}. Edge computing is different from the usual cloud computing approach. Instead of processing and storing data in centralized data centers far from users, edge computing involves positioning resources closer to users, specifically at the "edge" of the network. This means there are multiple computing nodes spread throughout the network, which reduces the burden on the central data center and makes data exchange much faster, as there is less delay in sending and receiving messages \cite{edge2}. Edge computing allows for the intelligent collection, analysis, computation, and processing of data at every IoT network edge. This implies that data can be filtered, processed, and used close to the devices or data sources, where it is generated. Edge computing makes everything faster and more effective by pushing smart services to the edge of the network. Making decisions and processing data locally can also help deal with significant limitations in networks and resources, and it can address concerns related to security and privacy too \cite{agriculture,ali2}.
\\Here is how this chapter is organized: Section 2 provides a comprehensive explanation of computing paradigms for IoT. Moving on to Section 3, a detailed introduction to edge computing paradigms is presented. Section 4 outlines the architecture of edge computing-based IoT. In Section 5, the focus shifts to illustrate the advantages of edge computing-based IoT. The enabling technologies for edge computing-based IoT are introduced in Section 6. Section 7, the chapter reviews edge computing in IoT-Based Intelligent Systems. Section 8 illustrates the challenges and future research directions for edge computing-based IoT. Finally, Section 9 concludes the chapter.
\section{Computing Paradigms for IoT}
\label{sec:2}
\begin{figure}
	\centering
	\includegraphics[width=0.7\textwidth]{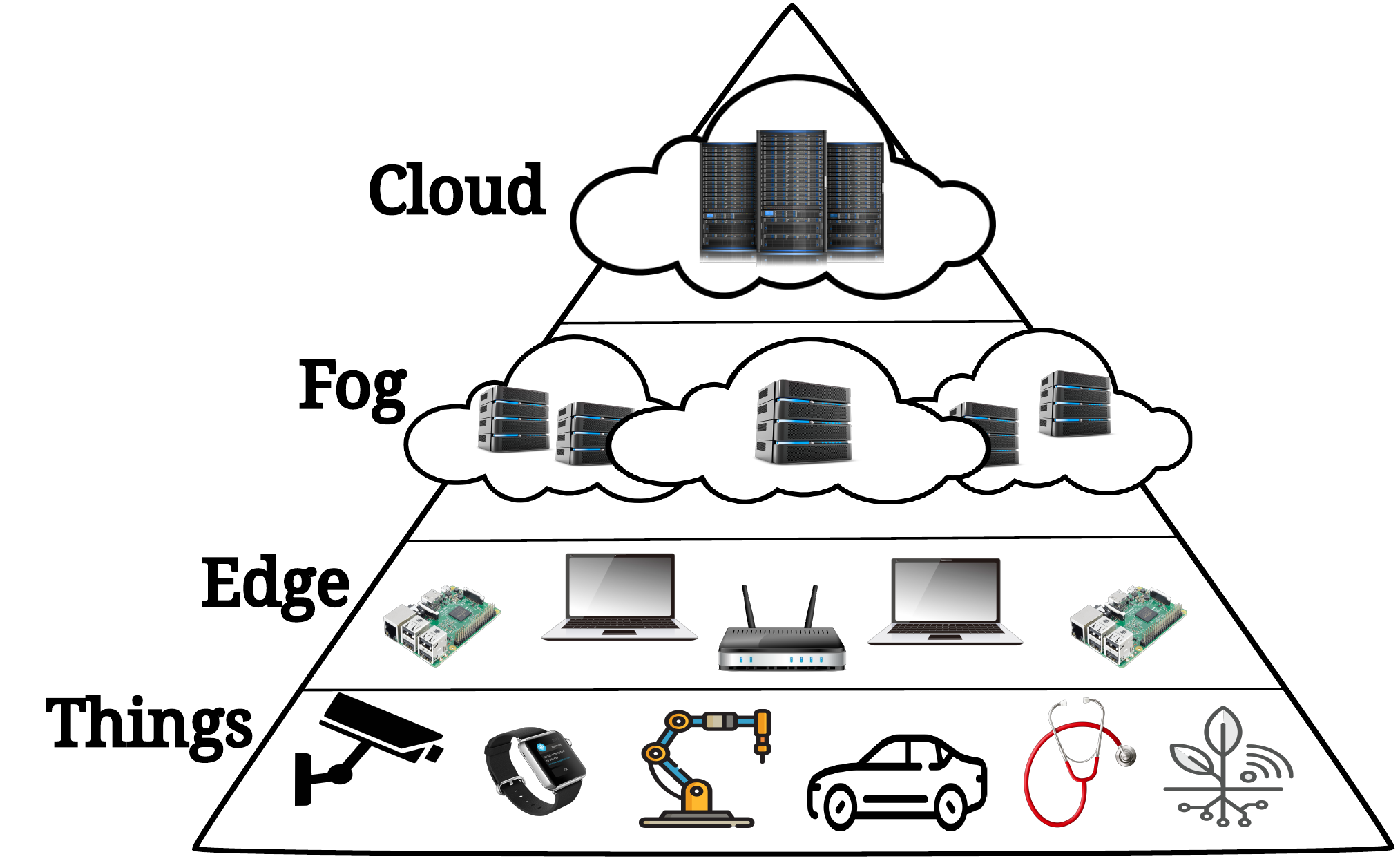}
	\caption{3-Tiers Architecture of Computing Paradigms}
	\label{fig:computing}
\end{figure}

This section describes the fundamental concepts underlying the three major computing paradigms and how they are integrated with IoT: cloud computing, edge computing, and fog computing, fgure 1 shows the architecture of the 3 tiers computing paradigms \ref{fig:computing}.
\subsection{Cloud Computing}
As mobile hardware evolves and improves, it will always face limitations in terms of available resources compared to stationary hardware. Regarding devices that people wear or carry for extended periods, prioritizing improvements in weight, size, and battery life takes precedence over enhancing computational power. This is a fundamental aspect of mobility rather than a transient restriction imposed by modern mobile electronics. Therefore, there will always be trade-offs when using computational power on mobile devices. The resource limitations of mobile devices can be solved simply and effectively by using cloud computing. With this approach, a mobile device can execute an application that requires a lot of resources on a robust remote server or a cluster of servers, allowing users to interact with the application through a lightweight client interface over the Internet \cite{cloudlet_3}.
\\The cloud computing paradigm gives end users on-demand services by utilizing a pool of computing resources. These resources include computing power, storage, and more, and they are all immediately available at any time \cite{cloud3}.
\\IoT and the cloud have had separate evolutionary processes. However, its integration has produced a number of benefits for both parties. On the one hand, IoT can greatly profit from the boundless capabilities of cloud computing to overcome its own technological limitations, such as storage, processing power, and energy requirements. The cloud can take advantage of IoT  through expanding its range of applications to handle real-world objects in a more distributed and adaptable fashion, thus providing new services in various real-life situations \cite{cloud1}.

\subsection{Edge Computing}
Typically, the architecture of the cloud is used to manage the massive amount of data that IoT devices produce. However, cloud computing encounters various challenges such as lengthy transmission time, increased bandwidth requirements, and latency between IoT devices and the cloud. The concept of edge computing has emerged to overcome these difficulties. This approach enhances scalability, latency, and privacy factors while enabling real-time predictions by processing data at the source \cite{edge_h1,ali9}.
\\ In an extension of cloud computing, edge computing places computer services at the edge of the network, where they are more accessible to end users. Edge computing shifts services, computational data, and applications out from cloud servers and toward the edge of a network.This enables content providers and application developers to provide users with services that are located nearby. Edge computing is unique in that it may be used for a variety of applications thanks to its high bandwidth, very low latency, and fast access to network data \cite{cloud3,ali10}.
\\ In the world of IoT, both Edge computing and Cloud computing offer major advantages due to their distinct characteristics, such as their capacity to execute complex computations and store large amounts of data. However, when it comes to IoT, edge computing outperforms cloud computing, despite having somewhat limited compute capability and storage capabilities. In particular, IoT demands fast responses rather than powerful computing and massive storage. Edge computing fulfills the requirements of IoT applications by offering satisfactory computing capability, sufficient storage, and fast response times. Edge computing, on the other hand, can also leverage IoT to expand its framework and adapt to the dynamic and distributed nature of edge computing nodes. These edge nodes can serve as providers and may consist of either IoT devices or devices with some residual computational capabilities \cite{edge2}.
\subsection{Fog Computing}
CISCO introduced the concept of fog computing in January 2014 \cite{fog2}. This computing paradigm offers numerous advantages across various fields, especially the IoT \cite{fog3,ali5}. According to Antunes, a senior official in charge of advancing corporate strategy at Cisco, edge computing is a division of fog computing. He explains that fog computing primarily focuses on managing the location of data generation and storage. In essence, edge computing involves processing data in proximity to its source of origin \cite{ali6}. Fog computing, on the other hand, leverages edge processing and the necessary network connections to transfer data from the edge to the endpoint\cite{fog2}. The fog computing system was not designed to replace cloud computing; instead, its development aimed to fill any service gaps present in cloud computing. Fog computing emphasizes on bringing abilities of cloud computing closer to the edge of the network so that users can access communication and software services faster. This approach works well for offering cloud solutions for highly mobile technologies like vehicular ad hoc networks (VANET) and the IoT \cite{fog1}. Fog computing serves the endpoints or edges of the network of interconnected devices. It prioritizes the analysis of time-sensitive data near the sources, sending only the selected and abridged data to the cloud \cite{fog2,ali7}.
\\ The concept of "fog as a service" (FaaS) is a new service possibility brought about by the integration of fog computing and IoT. According to this concept, a service provider creates a network of fog nodes throughout the area covered by its service, operating as a landlord to numerous tenants from diverse businesses. Each fog node provides storage, networking, and local computing capabilities. Through FaaS, customers can access services using innovative business models. Unlike clouds, which are often managed by huge businesses with sizable data centers, FaaS enables both small and large businesses to create and manage public or private computing, storage, and control services at different scales, meeting the requirements of various clients \cite{fog3,ali8}.

\section{Edge Computing Paradigms}
Edge computing emerged due to the evolution of cloud computing, and it provides different computing advantages. Several paradigms for operating at the edge of the network have been established throughout the growth of edge computing, including cloudlet and mobile edge computing. The two primary edge computing paradigms are introduced in this section.
\subsection{Cloudlet}
\label{subsec:3.1}
\begin{figure}
	\centering
	\includegraphics[width=0.5\textwidth]{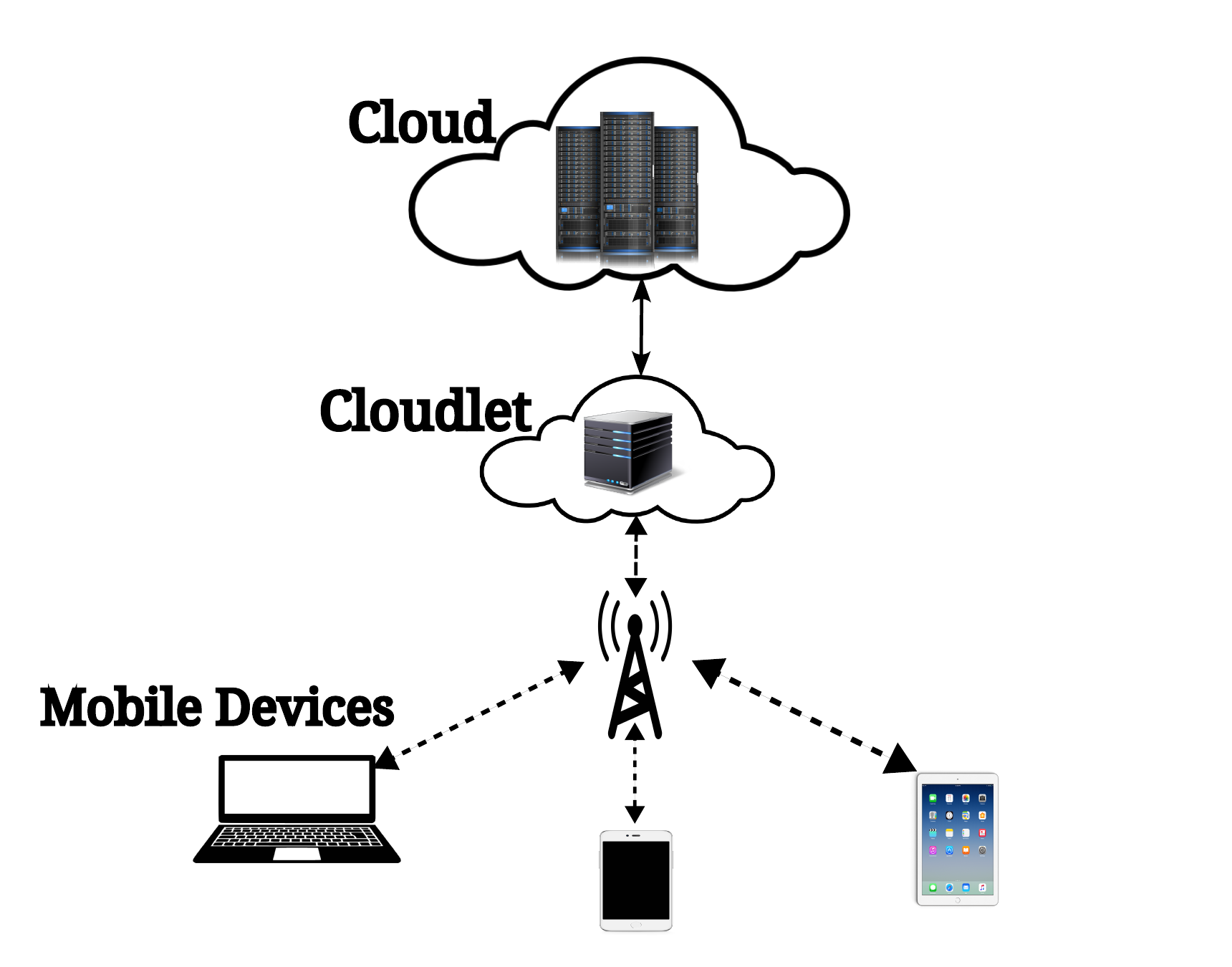}
	\caption{Cloudle Architecture }
	\label{fig:cloudlet}
\end{figure}

In 2009, Satyanarayanan and his team first proposed cloudlet computing as a remedy for the problems that can arise while using traditional cloud computing. These restrictions cover things like delay, jitter, and congestion \cite{cloudlet_3}. Cloudlets, as shown in figure \ref{fig:cloudlet}, are essentially small data centers,  often referred to as miniature clouds,which are frequently just a hop away from a user device ~\cite{cloudlet_1}. Instead of relying on a far-off "cloud," a nearby cloudlet with abundant resources can be used to alleviate the limited resources of a mobile device. To fulfill the requirement for real-time interactive responses, a solution is to establish a wireless connection with a nearby cloudlet that provides high-bandwidth, one-hop, and low-latency wireless access. In this situation, the mobile device serves as a lightweight client, with the cloudlet in close proximity handling the majority of the complex computational operations\cite{cloudlet_3}. Cloudlet relies on technologies like Wi-Fi, making it reliant on a strong internet connection\cite{MEC5}.
\\Owners of the network infrastructure, such as Nokia and AT \&  T, can enable cloudlets to be positioned in closer proximity to mobile devices, in hardware with smaller footprints than the cloud computing's massive data centers. As a result of their smaller footprint, cloudlets have less computational power than typical clouds, but they still have advantages over them including lower latency and energy usage ~\cite{cloudlet_1}.
\\A "data center in a box" is how cloudlets are like. It operates autonomously, with minimal power consumption, and only needs an Internet connection and access control for setup. This administrative simplicity correlates with a device-based computing architecture, enabling easy implementation in a variety of business premises such as doctor's offices or coffee shops. From an internal viewpoint,A cloudlet can be perceived as a cluster of computers equipped with multiple cores, fast internal connections, and a high-bandwidth wireless LAN. For safe implementation in unsupervised areas, the cloudlet can be housed in a protective casing designed to resist tampering with third-party remote monitoring of hardware integrity \cite{cloudlet_3}. To avoid any serious implications in the event of loss or malfunction, it is crucial to emphasize that the cloudlet should only store transient data and code, such as cached copies. In critical circumstances like military operations, disaster-affected areas, and even cyberattacks, cloudlets are essential. In these circumstances, the cloudlet, which is the middle layer in a 3-tier architecture consisting of the mobile, cloudlet, and cloud, is required to fill in and provide vital services, making up for the cloud's unavailability. Cloudlets also have the benefit of reducing the dangers linked to multi-hop networks, such as the possibility of DoS attacks\cite{cloudlet_2}.

\subsection{Mobile Edge Computing}
\begin{figure}
	\centering
	\includegraphics[width=0.4\textwidth]{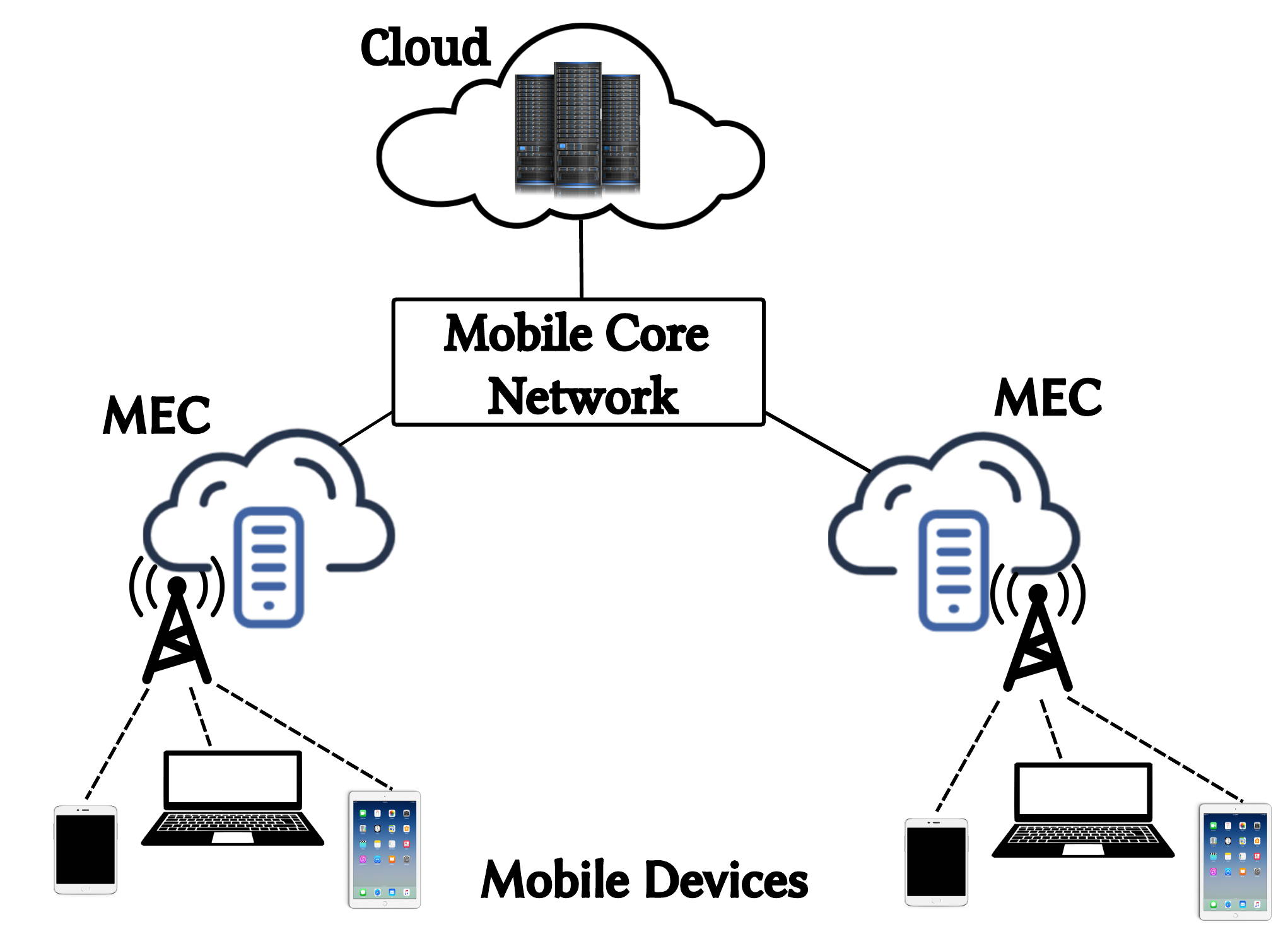}
	\caption{Mobile Edge Computing Architecture }
	\label{fig:mec}
\end{figure}

The idea behind a cloudlet is to strategically place powerful computers so that they can provide neighboring user equipment (UE) with computing and storage capabilities. Similar to WiFi hotspots, cloudlets deliver cloud services to mobile customers as opposed to offering internet connectivity. The fact that mobile UEs predominantly use WiFi connections to access cloudlets has a potential disadvantage because it requires users to alternate between the mobile network and WiFi anytime they need cloudlet services\cite{MEC1}. \\Another option that enables cloud computing at the edge is mobile edge computing (MEC), which first announced in 2014 by the European Telecommunications Standard Institute (ETSI). The MEC platform is characterized as a system that provides IT and cloud-computing functionalities within the Radio Access Network (RAN), positioned in close proximity to mobile subscribers \cite{MEC3}.
MEC is defined by the ETSI as having low latency, local processing and storage resources, network awareness, and better service quality supplied by mobile carriers. MEC makes mobile end-user services more accessible by providing compute and storage resources. These resources are intended to be deployed on mobile networks near end users. MEC resources can be used in a variety of locations, including radio access networks (RANs), indoor and outdoor base stations, and access points, which connect user equipment to Mobile Network Operators' (MNOs') core networks \cite{MEC4}.
\\Within the radio access network, MEC provides cloud computing capabilities, as depicted in Figure \ref{fig:mec}. Rather than routing mobile traffic from the core network to end-users, MEC creates a direct link between users and the nearest edge network empowered with cloud services. By deploying MEC at the base station, it enhances computation, mitigates bottlenecks, and reduces the risk of system failure \cite{MEC5}.
MEC is implemented on a virtualized platform that takes advantage of the most recent advancements in information-centric networks (ICN), network functions virtualization (NFV), and software-defined networks (SDN). A single-edge device with NFV at its core can provide computational services to numerous mobile devices by producing several virtual machines (VMs). These VMs can handle several tasks or perform diverse network operations at the same time \cite{MEC3}.

\section{Architecture of Edge Computing-Based IoT}
\begin{figure}
	\centering
	\includegraphics[width=0.5\textwidth]{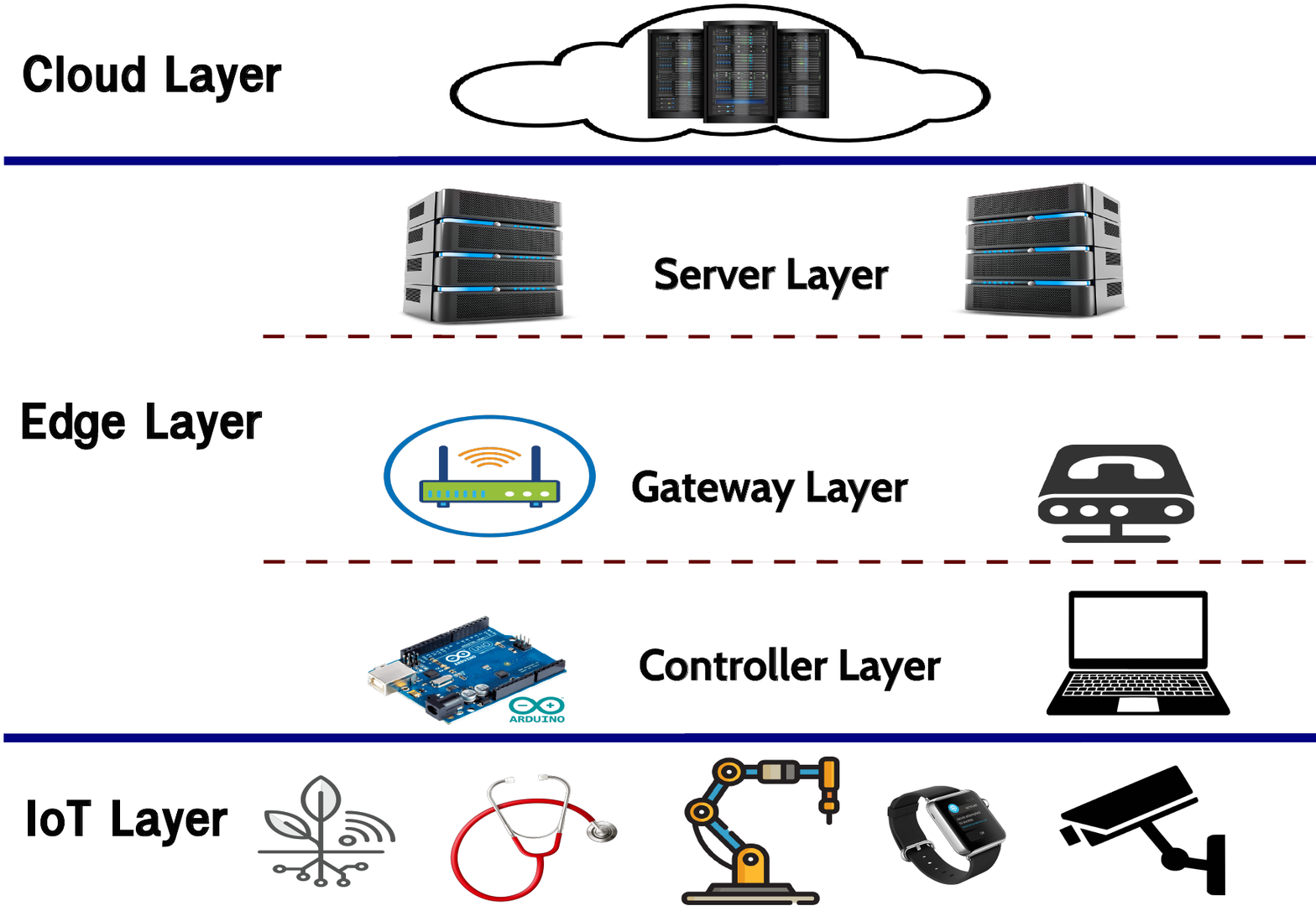}
	\caption{Edge Computing-based IoT Architecture }
	\label{fig:iot}
\end{figure}

In the context of IoT, edge computing is primarily focused on its implementation across various IoT scenarios, aiming to  minimize decision-making latency and network traffic\cite{advantages2}. The edge computing-based IoT architecture, as depicted in figure \ref{fig:iot}, consists of three distinct layers: IoT, edge, and cloud, all of these are built on top of existing edge computing reference designs. Our primary focus is to define the specific functions allocated to each layer and explore the communication mechanisms established among these layers\cite{advantages2,architecture2}.
\begin{enumerate}
	\item \textbf {IoT layer} : The IoT layer encompasses a broad spectrum of devices and equipment, such as smart cars, robots, smart machinery, handheld terminals, instruments and meters,  and other physical goods. These objects are tasked with overseeing the functioning of services, activities, or equipment. Furthermore, The IoT layer consists of actuators, sensors, controllers, and gateways constructed expressly for IoT contexts, which enable the administration of computational resources within IoT devices \cite{advantages2}.
	\item \textbf {Edge layer} : The main purpose of this layer is to receive, process, and send streams of data from the Device Layer. It offers real-time services like intelligent computing, security and privacy protection, and data analysis. Based on the equipment's ability to process data, Three further sub-layers are separated from the Edge Layer: the near-edge layer, the mid-edge layer, and the far-edge layer \cite{architecture2}.
		\begin{enumerate}[label=\alph*)]
			\item Far-Edge Layer (Edge controller layer) : In this layer, data is collected from the IoT layer by edge controllers and subsequently undergoes initial threshold assessment or data filtering. After that, the edge layer or cloud layer directs the control flow back to the IoT layer. After IoT device data has been collected, it is preprocessed to determine thresholds or perform data filtering. Consequently, the edge controllers in this layer must incorporate algorithm libraries tailored to the environment's configuration to consistently improve the strategy's efficiency. Additionally, these edge controllers should convey the control flow back to the IoT layer via the programmable logic controller (PLC) control or action control module after receiving decisions from the edge controller layer or upper layers \cite{advantages2}. 
			\item Mid-Edge Layer (Edge gateway layer) : This layer is often made up of edge gateways, which can connect to wired networks like industrial ethernet or wireless networks like 5G to receive data from the edge controller layer. Furthermore, the layer enables diverse processing capabilities and caches the accumulated data. Moreover,The edge gateways in this layer play a crucial role in shifting control from the upper layers, such as the cloud layer or edge server layer, to the edge controller layer. Simultaneously, they monitor the equipment in both the edge gateway layer and the edge controller layer\cite{advantages2}. The mid-edge layer has more storage and processing power than the far-edge layer, which can only carry out basic threshold judgment or data filtering. As a result, it can handle IoT layer data in a more thorough manner \cite{architecture2}.
			\item Near-Edge Layer (Edge server layer) : The edge server layer is equipped with robust edge servers. Within this layer, advanced and crucial data processing takes place. The edge servers leverage dedicated networks to gather data from the edge gateway layer and generate directional decision instructions based on this collected information. Additionally, platform administration and business application management features are anticipated for the edge servers in the edge server layer \cite{advantages2}.
		\end{enumerate} 
	\item \textbf {Cloud layer} : This layer primarily focuses on in-depth data mining and seeks to allocate resources optimally on a big scale, across a whole organization, a region, or even the entire country. Data from the edge layer is sent to the cloud layer through the use of the public network. Additionally, the edge layer has the ability to receive feedback from cloud layer-provided business applications, services, and model implementations \cite{advantages2}.
	
\end{enumerate}

\section{Advantages of Edge Computing-based IoT}
Edge computing plays a vital role as a computing paradigm for IoT devices, involving the utilization of cloud centers located near the IoT devices for tasks such as filtering, preprocessing, and aggregating IoT data \cite{advantages2}. The primary advantages of edge computing include:
\begin{enumerate}[label=\Alph*)]
	\item \textbf {Low Latency}: The close proximity and low latency of edge computing provide a solution to the response delay faced by User Equipment (UEs) while accessing typical cloud services. Edge computing can drastically reduce response time, which includes communication, processing, and propagation delays. Cloud computing typically results in an end-to-end latency of more than 80ms (or 160ms for response delay), making it unsuitable for time-sensitive applications such as remote surgery and virtual reality (VR), which require near-instantaneous replies within 1ms. Edge computing, on the other hand, benefits UEs by reducing total end-to-end delay and reaction delay due to their close proximity to edge servers. This enhancement enables faster and more efficient interactions for time-critical applications, meeting the requirements for tactile speed and responsiveness\cite{advantages}.
	\item \textbf {Energy Saving} : IoT devices often have limited energy supply due to their size and intended usage scenarios, yet they are expected to conduct complicated activities that are frequently power-intensive. It is difficult to design a cost-effective system to properly power numerous distributed IoT devices since regular battery charging or discharging is not always practicable or possible.However, edge computing offers a solution by enabling IoT devices to offload power-consuming computation tasks to edge servers. This not only substantially lowers energy use but also enhances processing efficiency, enabling billions of IoT devices to function optimally \cite{AI1}.
	\item \textbf {Security and Privacy} : among the most important features of cloud platform services is enhancing data security and privacy. Customers of these services can obtain centralized data security solutions from these providers, but any compromise of the centralizedly held data may have severe consequences. In contrast, edge computing has the benefit of allowing local deployment of customized security solutions. With this approach, less data transport is necessary because the majority of processing can be done at the network edge. As a result, there is a lower chance of data leakage during transmission, and less data is stored on the cloud platform, lowering the security and privacy risks \cite{advantages2}.
	\item \textbf {Location Awareness}: Edge servers with location awareness can acquire and handle data generated by User Equipment (UEs) based on their geographical locations. As a result, personalized and location-specific services can be offered to UEs, allowing edge servers to collect data directly from nearby sources without sending it to the cloud. This allows for more efficient and targeted service provisioning customized to specific UE needs\cite{advantages}.
	\item \textbf {Reduce Operational Expenses} : Transmitting data directly to the cloud platform incurs substantial operational expenses due to the demands for data transmission, sufficient bandwidth, and low latency. Edge computing, on the other hand, has the advantage of minimizing data uploading volume, resulting in less data transmission, lower bandwidth consumption, and lower latency. As a result, edge computing reduces operational costs when compared to direct data transfer to the cloud platform \cite{advantages2}.
	\item \textbf {Network Context Awareness}: Edge servers are able to understand the network context through network context awareness. This includes User Equipment (UE) information, such as allocated bandwidth and user locations, as well as real-time network conditions, such as traffic load in a network cell and radio access network specifics. With this invaluable knowledge, edge servers are better equipped to adapt and accommodate to the various UEs and network conditions, which leads to an optimum use of network resources. As a result, edge servers can effectively handle a large amount of traffic, improving network performance. Additionally, the availability of fine-grained information enables the development of services that are specifically customized to the needs of various traffic flows and individual users \cite{advantages}.
	
\end{enumerate}

\section{Enabling Edge computing-based IoT Technologies}
Edge computing-based IoT can be implemented with the integration of several enabling technologies. This section illustrates the relevant enabling technologies by using artificial intelligence and lightweight virtualization as examples.
\subsection{Edge Intelligence}
As the need for intelligent edge devices has grown, the industry has responded with innovation and the adoption of intelligent edge architectures. These innovative architectures support real-time, mission-critical applications that work with a wide variety of devices. Any machine can qualify as intelligent if it mimics human behaviors and skills including perception, attention, thinking, and decision-making. Machine learning has gained a lot of traction as a field of advancement in artificial intelligence. This has led to a surge in the presence of intelligent devices, fueled primarily by advancements in deep learning techniques \cite{edge_h1}. 
\\ Deep Neural Networks (DNNs) have received substantial attention in the Machine Learning era because of their unrivaled performance across different use cases such as computer vision, natural language processing, and image processing \cite{AI3}. 
Notably, deep learning has even outperformed human players in complex games like Atari Games and the game of Go. The integration of deep learning and edge computing holds promise for addressing challenges and opening up new possibilities for applications.  On one hand, edge computing applications greatly benefit from the powerful processing capabilities of deep learning, enabling them to handle intricate scenarios like video analytics and transportation control. On the other hand, edge computing offers specialized hardware foundations and platforms, such as the lightweight Nvidia Jetson TX2 development kit, to effectively support deep learning operations at the edge\cite{AI1}.
Many techniques have been introduced to improve the performance of deep learning when performed on edge computing devices, such as :
\begin{enumerate}[label=\Alph*)]
	\item \textbf {Model design}: When machine learning researchers design DNN models for resource-constrained devices, they commonly emphasize creating models with fewer parameters in order to minimize memory usage and execution latency, while still maintaining high accuracy. Several techniques are employed to achieve this, including MobileNets, SSD, YOLO, and SqueezeNet. These methods are aimed at optimizing DNN models for efficient performance on such devices\cite{AI2}.
	\item \textbf {Run-Time Optimizations}: Depending on the particular requirements of the application, suitable run-time optimizations can be employed to minimize the quantity of samples that need to undergo processing. For instance, in object detection applications, a high-resolution image can be divided into smaller images (tiling), and a selection criterion can be used to choose images with high activity regions. This approach allows the design of DNNs that can handle smaller inputs, resulting in improved computational and latency efficiency.
	\item \textbf {Hardware}:In the pursuit of accelerating deep learning inference, hardware manufacturers are adopting various strategies. These include utilizing already existing hardware like CPUs and GPUs, as well as developing custom Application-Specific Integrated Circuits (ASICs) dedicated to deep learning tasks, like Google's Tensor Processing Unit (TPU). Additionally, there are novel custom ASICs like ShiDianNao, which prioritize efficient memory access to minimize latency and energy consumption. FPGA-based DNN accelerators also show promise, as FPGAs can deliver fast computation while remaining reconfigurable\cite{AI2}.
	
\end{enumerate}
\subsection{Lightweight Virtualization}
\begin{figure}
	\centering
	\includegraphics[width=0.7\textwidth]{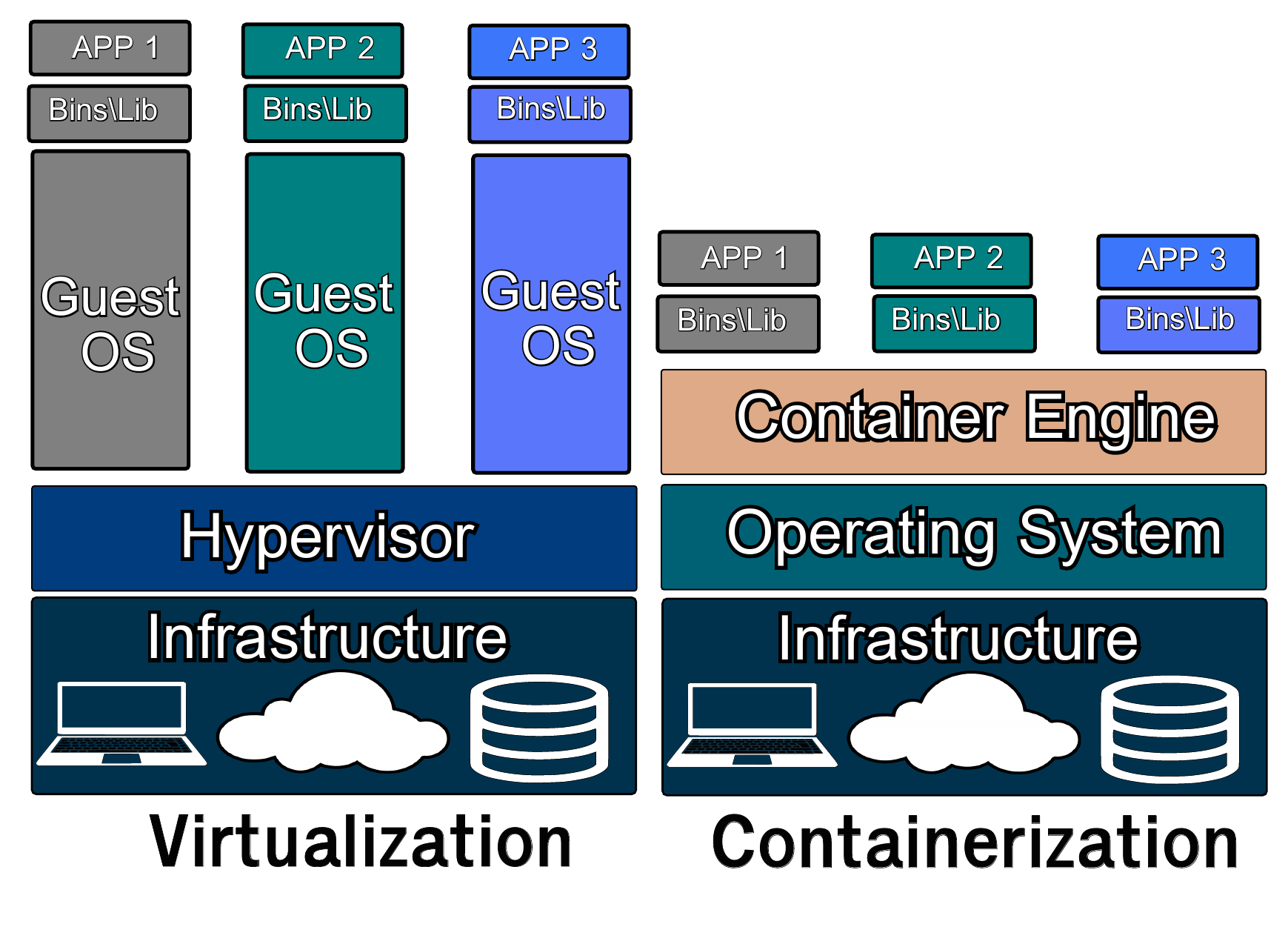}
	\caption{ Virtualization vs Containerization }
	\label{fig:container}
\end{figure}

Virtualization technologies are widely employed in cloud computing due to their effective method of harnessing the cloud's capabilities by partitioning a physical host into smaller, more manageable virtual components. By leveraging these technologies, cloud computing services become more user-friendly and economically efficient. Hypervisors such as VirtualBox and VMware are frequently used in cloud computing hardware virtualization. However, this approach has limitations such as increased resource cost, longer startup times, and larger attack surfaces. To solve these limitations, lightweight virtualization technologies such as Unikernels and Containers have evolved and are currently used in both cloud and edge computing. These lightweight virtualization technologies offer fast deployment and high efficiency, effectively overcoming the limitations posed by traditional hypervisor-based virtualization\cite{container1}.Considering that the computational capabilities of edge computing devices are less potent than data centers, the adoption of emerging lightweight virtualization technologies offers numerous advantages. These benefits encompass swift initialization, minimal overhead, high instance density, and commendable energy efficiency, making them well-suited for the edge computing environment \cite{continer2}.
\\ Lightweight virtualization technology is critical in edge computing because it allows the deployment of resource management, orchestration, and isolation services without the need to account for different hardware configurations. This technology has brought about a significant transformation in software development and deployment practices. Container-based virtualization can be regarded as a lightweight alternative to the traditional Hypervisor-based virtualization \cite{container1}. Container-based virtualization offers a different level of abstraction in terms of isolation and virtualization when compared to hypervisors, as illustrated in figure\ref{fig:container}. Hypervisors virtualize hardware and device drivers, resulting in increased overhead. Containers, on the other hand, isolate processes at the OS level \cite{continer2}. Containers allow independent applications to be isolated with their own virtual network interfaces, process spaces, and file systems because they share the same host machine's operating system kernel.  Containers allow for a higher number of virtualized instances with lower image volumes, all executing on a single machine thanks to the shared kernel feature \cite{container1}.

\section{Edge Computing in IoT-Based Intelligent Systems:Case Studies}
With the rise of intelligent systems, edge computing offers the most efficient computing and storage solutions for devices with limited computational capabilities.  This section delves into the applications of edge computing in IoT-based intelligent systems, including healthcare, manufacturing, agriculture, and transportation. We chose these four case studies because they have a substantial impact on improving human life.
\subsection{Edge Computing in IoT-Based Healthcare}
The term "geriatric care" refers to a branch of healthcare that emphasizes meeting the special mental, physical, and social needs of the aged. Geriatric care, which is specifically designed to meet the unique demands of elderly individuals, seeks to enhance their general well-being and health while successfully treating age-related illnesses and diseases. Its ultimate goal is to give them the means to maintain their independence, preserve their well-being, and enjoy the greatest degree of comfort as they age \cite{edge_h2}. In the area of "Geriatric Care," the danger of falling is regarded as a crucial concern. Unfortunately, many older people fall and unfortunately pass away because they lack good balance. While the fall itself may be the primary cause, the severity of the outcome stems from the inability to recover, leading to deteriorating physical and cognitive health. Numerous studies back up the idea that elderly people can potentially avoid physical consequences like brain injuries if immediate aid is given within seven minutes of a fall. As a result, the death and disease rates among the aging population would both dramatically decline. In \cite{edge_h1} the authors suggest an intelligent edge-monitoring system that utilizes cameras to detect instances of falling. Real-time video analysis is essential for continuously capturing photos, classifying them as either normal (sleeping) or abnormal (falling) circumstances, and instantly sounding an alarm in emergency scenarios. Due to the significant amount of data involved, relying solely on cloud processing would be impractical because of the resulting delays. As a result, cameras serve as IoT devices, gathering data and sending it to nearby edge computing equipment for processing locally. In this approach, the Edge computing server is equipped with Deep Learning models that have already undergone pre-training and are specifically created to detect falls with high accuracy and low latency. They successfully reduce transmission delays by maintaining the computing process' independence from the cloud. This approach not only allows them to generate valuable insights on-site, reducing response time and latency but also addresses privacy concerns by handling sensitive and personal data at the edge.

\subsection{Edge Computing in IoT-Based Manufacturing}
Regarding intelligent manufacturing, the growing number of terminal network devices has presented new issues in terms of data center operation and maintenance, scalability, and dependability. To tackle these challenges, edge computing has advanced, which moves computation from centralized data centers to the network's edge. This approach enables intelligent services to be deployed near the manufacturing units, meeting essential demands such as highly responsive cloud services, data analytics via edge nodes,  and a privacy-policy plan \cite{manufacturing1}. In \cite{manufactoring2}, the authors introduced a cutting-edge visual sorting method created especially for adaptable manufacturing systems. The system utilizes a visual sorting approach based on CNNs. To support this system, they developed a cloud-edge computing platform, which makes it easier to quickly compute and continuously maintain and improve services.
\subsection{Edge Computing in IoT-Based Agricultural }
The IoT is typically applied in agricultural development through a monitoring network comprised of a considerable number of sensor nodes, As a result, agriculture increasingly moves away from a production model that is centered on humans and toward one that is centered on information and software \cite{agriculture}. Concerning the notion of edge computing in the context of Agricultural IoT, numerous researchers have made contributions from diverse perspectives. The authors of \cite{agriculture2} developed a system that meets the demanding needs of Precision Agriculture (PA) by integrating automation, IoT technologies, edge and cloud computing through virtualization . Three essential layers make up a multi-tier platform that has been developed: (1) a local layer of Cyber-Physical Systems (CPS) that is connected to agricultural greenhouses. (2)The authors suggest a new edge computing architecture in which control modules are placed on virtualized nodes near the access network. (3) A cloud section outfitted with powerful computing and data analytics tools to help farmers make smart crop management decisions. The entire system was successfully tested in a real greenhouse located in southeast Spain. Through specialized software that was accessible to the end farmers via the platform, this innovation made it possible to control a closed hydroponic system in real-time. To validate the effectiveness of the architecture, two tomato crop cycles were conducted. The results showed remarkable benefits compared to a traditional open crop approach. Significant water savings of over 30\% were achieved, which is particularly crucial in their semi-arid region. Additionally, certain nutrients saw improvements of up to 80\% thanks to the system's efficient management.
\subsection{Edge Computing in IoT-Based Transportation}
The Internet of Vehicles (IoV), a new paradigm introduced by the IoT, employs edge computing to offer ground-breaking applications for transportation systems. Using sensors and geofencing technologies, IoV connects various cars with Roadside Units (RSUs) and other vehicles in an Intelligent Transportation System (ITS). Edge cloudlets are used by IoV for service provisioning and orchestration. Currently, substantial research on smart vehicles is being undertaken in both academic and industrial domains \cite{IoV1}. Traffic flow detection plays a vital role in ITS. By obtaining real-time urban road traffic flow data, ITS can intelligently guide measures to alleviate traffic congestion and reduce environmental pollution. In\cite{IoV2}, the YOLOv3 (You Only Look Once) model was used by the authors to create a vehicle detecting method. The YOLOv3 model was trained on an extensive dataset of traffic data and subsequently pruned to achieve optimal performance on edge devices. Additionally, by retraining the feature extractor, they improved the DeepSORT (Deep Simple Online and Realtime Tracking) algorithm, enabling multi-object vehicle tracking. Through the integration of vehicle detection and tracking algorithms, they developed a counter for real-time vehicle tracking capable of accurately detecting traffic flow. Finally, the Jetson TX2 edge device platform received and implemented the vehicle detection network and multi-object tracking network.

\section{Challenges and Future research Directions }
In the preceding section, we outlined four possible uses of edge computing in IoT-based systems. To achieve the full potential of IoT, we emphasize the need for seamless collaboration between IoT devices and edge computing.
Now, in this section, we will summarize some of the challenges faced in implementing edge computing in IoT-based systems and propose potential solutions and research opportunities. These include resource allocation, heterogeneity, privacy and security, and microservices.

\begin{itemize}
	\item \textbf {Resource Allocation } : Edge devices play an essential role in enabling latency-critical services. The majority of end IoT devices often experience resource limitations; for example, local CPU computation capabilities and battery capacity are frequently constrained. Some workloads can be offloaded to more powerful edge devices to bypass these constraints and meet the performance requirements of applications. Edge computing improves IoT device capabilities, allowing them to handle more resource-intensive applications. However, Practically speaking, edge computing devices have a finite amount of processing power. As a result, it is unable to handle the massive computing tasks generated by all of the end devices in its service region. As a result, the allocation of resources becomes highly crucial in such environments. Traditional optimization approaches like convex optimization and Lyapunov optimization have been used to tackle the computation offloading problem and find the best scheme. However, these methods are limited when it comes to making optimal decisions in dynamic environments. In contrast, modern resource allocation algorithms, powered by artificial intelligence and deep learning, such as deep reinforcement learning, offer more effective solutions for achieving optimal allocation.
	\item \textbf {Heterogeneity } : The existence of numerous computing technologies in edge computing, such as distinct hardware architectures and operating systems, has made it difficult to develop a viable approach that can be employed with ease in diverse scenarios. This problem can be solved by developing a programming model for edge nodes using software-based techniques, which enables workloads to be executed effectively on numerous hardware configurations simultaneously.
	\item \textbf {Privacy and Security} :At the network's edge, the primary services that require assurance are the protection of usage privacy and data security. When IoT is used in a home, the usage data collected can be used to infer important private information. For instance, examining the patterns of energy or water use can show whether the home is vacant or occupied. This presents a significant challenge in providing services while safeguarding privacy. Having a reliable architecture is crucial before users can feel confident in embracing new technologies. "Privacy by design"  can be considered as a reliable approach to enhance security in edge computing. It involves incorporating privacy features directly into the design, taking preventive measures instead of just reacting after privacy breaches, and ensuring data privacy throughout its entire lifecycle.
	\item \textbf {Microservices} : Recently, both edge and cloud services have been changing from monolithic, standalone systems to loosely coupled, independent microservices. When running complex computations such as deep learning, there are several software requirements, and it's important to find a way to separate different deep learning services when using shared resources. Currently, the microservice framework, which can be used to host complex services on the edge, is still developing and in its early stages. However, it shows great potential for efficiently introducing services in the future.
\end{itemize}

\section{Conclusion}
As IoT continues to grow, edge computing is increasingly regarded as a promising and viable solution to address the complexities of managing numerous sensors and devices, along with the demands for resources they require. Edge computing, in contrast to standard cloud computing, involves placing data processing and storage to the edge of the network, bringing them closer to the end users. Thus, by dispersing compute nodes across the network, it is possible to reduce message exchange latency and relieve the computational load on the centralized data center. In conclusion, our chapter has explored the computing paradigms for IoT, edge computing paradigms like cloudlet and MEC, the architecture of edge computing-based IoT, the benefits it offers, and the enabling technologies such as artificial intelligence and lightweight virtualization. Additionally, it presented case studies showcasing how edge computing is applied in intelligent systems based on IoT and highlighted the issues with current research and suggested future directions for further exploration in this field.

\end{document}